\documentclass{jps-cp}
\usepackage{txfonts} 
\usepackage{braket}
\usepackage{overcirc}
\newcommand{\Hc}{{\cal H}}
\newcommand{\Qhat}{\hat{Q}}
\newcommand{\Phat}{\hat{P}}
\newcommand{\Hhat}{\hat{H}}

\title{Nuclear reaction path and requantization of TDDFT}

\author{Kai \textsc{Wen}$^{1,2}$ and Takashi \textsc{Nakatsukasa}$^{2,3,4}$}

\inst{$^{1}$University of Surrey, Guildford, UK \\
$^{2}$Center for Computational Sciences, University of Tsukuba,
Tsukuba 305-8577, Japan\\
$^{3}$Faculty of Pure and Applied Sciences, University of Tsukuba,
Tsukuba 305-8577, Japan\\
$^{4}$iTHES, RIKEN, Wako 351-0198, Japan
}

\recdate{January 5, 2018}

\abst{
Using a theory of large amplitude collective motion,
the adiabatic self-consistent collective coordinate method,
we derive reaction path for the fusion process at sub-barrier energies.
The collective Hamiltonian to describe the fusion process is
constructed, based on the obtained reaction path and canonical variables.
We study the reaction of $N=Z$ stable nuclei,
$\alpha+^{16}$O, $^{16}$O$+^{16}$O, and $\alpha+^{12}$C.
The results suggest that, after two nuclei touch,
the reaction path is significantly deviated from the simple relative motion,
which may affect the deep sub-barrier fusion cross section.
}

\kword{sub-barrier fusion, collective motion,
ASCC, TDHF, energy density functional theory}

\begin{document}
\maketitle

\section{Introduction}

It has been a long-standing problem in nuclear theory to give a non-empirical
description of low-energy nuclear reaction, starting from a
quantum many-body Hamiltonian.
Recently, we see significant developments in
the real-time simulation using
time-dependent Hartree-Fock (TDHF) and time-dependent Hartree-Fock-Bogoliubov
(TDHFB) theories \cite{Sim12,NMMY16,SY16,UOS16,BMRS16,MSW17,Sek17,USY17}.
However, the real-time simulations are not applicable to
low-energy quantum reactions, such as
sub-barrier fusion reaction and spontaneous fission.
This is due to a lack of full quantum nature in the TDHF(B) dynamics.
To recover the missing quantum fluctuation in particular modes,
the requantization of the TDHF(B) dynamics is a useful idea
\cite{Neg82,NMMY16}.

The first major task is to identify specific degrees of
freedom to quantize.
This is not trivial for low-energy collective motion in nuclear physics.
In order to achieve this goal, we use a method based on the
adiabatic self-consistent collective coordinate (ASCC) method
\cite{MNM00,HNMM07,HNMM09,Nak12,NMMY16}.
The ASCC method consists of a set of equations for collective submanifold
associated with ``slow'' collective motion,
decoupled from intrinsic degrees of freedom of ``fast'' motion.
The solution of those equations provides a series of time-even
Slater determinants
which are located on the collective submanifold.
In this manner, we are able to extract the collective degrees of freedom
for requantization.

The next task is to perform the quantization.
Since we obtain a series of Slater determinants to span the collective
subspace, we may consider superposition of those states to take into
account the quantum fluctuation associated with the collective motion.
The idea, which is similar to the generator coordinate method (GCM)
\cite{RS80}, sounds attractive.
However, starting from an effective Hamiltonian with density dependence,
it is not as straightforward as we may think
\cite{AER01,DSNR07,DBBLL09}.
In general, what we start from is the energy density functional (EDF),
instead of the Hamiltonian,
then, the Hamiltonian kernel for the GCM is not well defined.
Furthermore, there is no guarantee that dynamical effects are properly
taken into account by superposing only time-even Slater determinants.
In the case of center-of-mass motion in the GCM,
the use of complex generator coordinates is necessary to reproduce
the correct total mass,
implying that a pair of collective coordinate and momenta
should be simultaneously treated as generator coordinates
\cite{PT62,RS80}.

In the present paper, we adopt another approach, namely,
construction of the collective Hamiltonian with a few collective variables.
The ASCC method provides us with generators of canonical collective variables
$(q,p)$ and collective inertial masses,
in addition to the time-even Slater determinants parameterized as
$\ket{\Phi(q,p)}$.
The inertial mass in the ASCC method takes into account effect of
the time-odd mean fields.
Therefore, it reproduces the exact total mass for the center-of-mass motion.
Utilizing this procedure, we may construct the collective Hamiltonian
to describe the low-energy nuclear reaction.
Since it is given in terms of the small number of canonical variables,
the quantization is rather easy.
We adopt the standard Pauli's prescription for the canonical quantization.

This paper is organized as follows.
In section \ref{sec:SCC}, we give basic equations of the ASCC method.
In section \ref{sec:numerical}, some details of the numerical procedure
are shown.
Then, the result for the nuclear fusion reaction at sub-barrier energies
is presented in section \ref{sec:result}.
Finally, the summary and the perspectives are given in section \ref{sec:summary}.

%
%
%

\section{Self-consistent collective coordinate method}
\label{sec:SCC}

\subsection{Basic principles}

Marumori and coworkers proposed the self-consistent collective coordinate
(SCC) method, in order to formulate a theory for the large amplitude
collective motion
without assuming adiabaticity \cite{MMSK80}.
In the original formulation of the SCC method,
collective coordinates and collective momenta are equally treated,
then, basic equations determining the collective submanifold
are derived from a condition of maximal decoupling of the collective motion
from other non-collective degrees of freedom.
In this subsection, we recapitulate the basic principles of the SCC method.

Let us consider a TDHF state $\ket{\phi(t)}$ describing
a collective motion.
If this collective motion is decoupled from other intrinsic degrees of
freedom, 
the state $\ket{\phi(t)}$ is confined in a ``collective submanifold''
parameterized by $(q^i,p_i)$ with $i=1,\cdots, K$,
embedded in the full TDHF phase space parameterized by
$(\xi^\alpha,\pi_\alpha)$ with $\alpha=1,\cdots, M$
($K\ll M$).
The time evolution of $\ket{\phi(t)}$ is governed by
a small number of collective coordinates
$q(t)$  and collective momenta $p(t)$

The SCC method are based on the
``invariance principle of the TDHFB equation'', which
requires that the TDHF equation of motion is invariant
in the collective submanifold.
\begin{equation}
\delta\bra{\phi(q,p)} \left\{ i\frac{\partial}{\partial t} - \hat{H} \right\} \ket{\phi(q,p)}=0.
\label{invariance principle}
\end{equation}
Here, the variation $\delta$ is given by
all possible particle-hole excitations with respect to the
state $\ket{\phi(q,p)}$.
In the first sight, this looks the same as
the well-known time-dependent variational principle.
In fact, if we take ``tangential'' variations on the collective submanifold,
we obtain the Hamilton equations of motion
\begin{equation}
\frac{dq^i}{dt}=\frac{\partial \Hc}{\partial p_i}
,\quad
\frac{dp_i}{dt}=-\frac{\partial \Hc}{\partial q^i},
\label{Hamilton_eq}
\end{equation}
where the total energy $\Hc(q,p)\equiv \bra{\phi(q,p)} \hat{H} \ket{\phi(q,p)}$
plays the role of the classical collective Hamiltonian.
This determines the time dependence of $q(t)$ and $p(t)$.

On the other hand, for Eq. (\ref{invariance principle}),
variations ``normal'' to the submanifold determine
states $\ket{\phi(q,p)}$, namely a decoupled collective submanifold itself.
Equation~(\ref{invariance principle}) can be rewritten as
\begin{equation}
\delta\bra{\phi(q,p)} \left\{
\hat{H} -  \frac{\partial \Hc}{\partial p_i}  \overcirc{P}_i(q,p)
- \frac{\partial \Hc}{\partial q^i}\overcirc{Q}^i(q,p)  \right\}
 \ket{\phi(q,p)}=0 ,
\label{eq of submanifold1}
\end{equation}
in terms of the local infinitesimal generators defined by
\begin{eqnarray}
\overcirc{P}_i(q,p) \ket{\phi(q,p)} &=&   i \frac{\partial}{\partial q^i} \ket{\phi(q,p)},
\label{generator_P}
\\
\overcirc{Q}^i(q,p) \ket{\phi(q,p)} &=& -i \frac{\partial}{\partial p_i} \ket{\phi(q,p)},
\label{generator_Q}
\end{eqnarray}
which are one-body operators.
We call (\ref{eq of submanifold1}) 
``equation of collective submanifold''.
The canonicity condition, that $(q,p)$ are the canonical variables,
can be written as
\begin{eqnarray}
\bra{\phi(q,p)} \overcirc{P}_i(q,p) \ket{\phi(q,p)} &=& p_i + \frac{\partial S}{\partial q^i},
\label{canonicity conditions1}
\\
\bra{\phi(q,p)} \overcirc{Q}^i(q,p) \ket{\phi(q,p)} &=& -\frac{\partial S}{\partial p_i},
\label{canonicity conditions2}
\end{eqnarray}
where $S(q,p)$ is an arbitrary differentiable function of $q$ and $p$
\cite{MMSK80,YK87}.

\subsection{Expansion with respect to collective momenta}

In this paper, we study the sub-barrier fusion reaction.
At least, near the touching point, we can assume
that the relative motion of two nuclei are much slower than
the intrinsic single-particle motion.
Thus, we may take a kind of adiabatic limit, using the expansion with
respect to the collective momenta $p$.
It is called ``adiabatic SCC (ASCC) method'' \cite{MNM00}.

Now, we define the state at the limit of $p\rightarrow 0$,
$\ket{\phi(q,p)}\rightarrow \ket{\phi(q)}$.
Then, we express the TDHF state $\ket{\phi(q,p)}$ as
\begin{equation}
 \ket{\phi(q,p)}  =  \exp\left\{ i p_i \Qhat^i(q) \right\}
 \ket{\phi(q)} ,
\label{eq:ASCCstate}
\end{equation}
where $\Qhat^i(q)$ are one-body operators corresponding to
infinitesimal generators of $p_i$ locally defined at the state $\ket{\phi(q)}$,
namely, $\Qhat^i(q)\ket{\phi(q)}=\overcirc{Q}^i(q,p=0)\ket{\phi(q)}$.
Similarly, we may define the generators of $q^i$ at $\ket{\phi(q)}$ as
$\Phat^i(q)\ket{\phi(q)}=\overcirc{P}^i(q,p=0)\ket{\phi(q)}$.

In the adiabatic limit, it is convenient to adopt the function,
$S(q,p)=\mbox{constant}$.
This fixes the canonical coordinate systems within the ambiguity of
point transformation \cite{YK87}.
Under the point transformation, the quadratic form of the kinetic
energy terms are invariant.
Thus, we expand the classical collective Hamiltonian up to the
second order in momenta.
\begin{eqnarray}
&&\Hc(q,p)= V(q) + \frac{1}{2} B^{ij}(q) p_i p_j , \\ 
&&V(q) = \Hc(q,p=0) 
,\quad
B^{ij}(q)=\left.\frac{\partial^2 \Hc}{\partial p_i \partial p_j}\right|_{p=0} .
\nonumber
\end{eqnarray}
The collective inertial tensors $B_{ij}(q)$ are defined as the inverse matrix
of $B^{ij}(q)$, ~$B^{ij} B_{jk}=\delta^i_k$.
With this choice of the function $S$,
the canonicity conditions (\ref{canonicity conditions1}) and
(\ref{canonicity conditions2}) lead to
the weak canonical variable condition,
\begin{equation*}
\bra{\phi(q)} \left[ \Qhat^i(q), \Phat_j(q) \right] \ket{\phi(q)} = i\delta_{ij}. 
\end{equation*}

From the equation of collective submanifold (\ref{eq of submanifold1}),
we may derive the following three equations,
as the zeroth, first and second order in momenta,
respectively \cite{MNM00,NMMY16}.
\begin{eqnarray}
\delta\bra{\phi(q)}\Hhat_M(q)\ket{\phi(q)} = 0,&&
\label{eq:mfHFB}
\\
\delta\bra{\phi(q)} \left[\Hhat_M(q), \Qhat^i(q)\right]
- \frac{1}{i} B^{ij}(q) \Phat_j(q)
+\frac{1}{2}\left[\frac{\partial V}{\partial q^j}\Qhat^j(q),
 \Qhat^i(q)\right]
\ket{\phi(q)} = 0, &&
\label{eq:ASCC1}
\\
\delta\bra{\phi(q)} \left[\Hhat_M(q),
 \frac{1}{i}\Phat_i(q)\right] - C_{ij}(q) \Qhat^j(q)
- \frac{1}{2}\left[\left[\Hhat_M(q), \frac{\partial V}{\partial
 q^k}\Qhat^k(q)\right], B_{ij}(q) \Qhat^j(q)\right]
\ket{\phi(q)} = 0. &&
\label{eq:ASCC2}
\end{eqnarray}
Here, $\Hhat_M(q)$ represents the moving Hamiltonian
\begin{equation}
  \Hhat_M(q) = \Hhat 
 -  \frac{\partial V}{\partial q^i}\Qhat^i(q) ,
\end{equation}
and 
\begin{eqnarray}
 C_{ij}(q) &=& \frac{\partial^2 V}{\partial q^i \partial q^j} - \Gamma_{ij}^k\frac{\partial V}{\partial q^k}, \\
 \Gamma_{ij}^k(q) &=& \frac{1}{2} B^{kl}\left( \frac{\partial B_{li}}{\partial q^j}
 + \frac{\partial B_{lj}}{\partial q^i} - \frac{\partial B_{ij}}{\partial q^l} \right).
\end{eqnarray}
For Eq. (\ref{eq:ASCC2}), we use the the derivative of Eq. (\ref{eq:mfHFB})
with respect to $q^j$ to eliminate $d\Qhat^i/dq^j$.
These equations are the basic equations for the ASCC method.

\subsection{One-dimensional case and collective Hamiltonian}

In the case of one-dimensional ($K=1$) collective path,
the last term in Eq. (\ref{eq:ASCC1}) vanishes.
In the present paper, 
we also neglect the last term in Eq. (\ref{eq:ASCC2})
for simplicity.
Then, Eqs. (\ref{eq:ASCC1}) and (\ref{eq:ASCC2}) are simplified as
\begin{eqnarray}
\delta\bra{\phi(q)} \left[\Hhat_M(q), \Qhat(q)\right]
+ i B(q) \Phat(q)
\ket{\phi(q)} = 0, &&
\label{eq:ASCC1-2}
\\
\delta\bra{\phi(q)} \left[\Hhat_M(q),
 i \Phat(q)\right] + C(q) \Qhat(q)
\ket{\phi(q)} = 0. &&
\label{eq:ASCC2-2}
\end{eqnarray}
Here, we omit the indices, $i=j=1$.
Solving these equations, we determine the microscopic structure of
the generators, $\Qhat(q)$ and $\Phat(q)$.
Then, the zeroth-order equation (\ref{eq:mfHFB})
determines the state $\ket{\phi(q)}$.
Since $\Hhat_M(q)$ contains the operator $\Qhat(q)$,
the solution should be self-consistent,
simultaneously satisfying Eqs. (\ref{eq:mfHFB}),
(\ref{eq:ASCC1-2}), and (\ref{eq:ASCC2-2}).

The solutions of Eqs. (\ref{eq:ASCC1-2}) and (\ref{eq:ASCC2-2}) 
provide us not only the local generators $\Qhat(q)$ and $\Phat(q)$,
but also the collective inertial mass $B(q)$ and the curvature $C(q)$.
Using the obtained inertial mass, we may write the collective Hamiltonian
as
\begin{equation}
H_c(q,p)=\frac{1}{2}B(q)p^2 + V(q) ,
\label{collective_Hamiltonian}
\end{equation}
where the potential is given by
$V(q)\equiv \bra{\phi(q)}\Hhat\ket{\phi(q)}$.

\section{Numerical procedure}
\label{sec:numerical}

In this paper, we study low-energy fusion reaction at sub-barrier
energies.
We employ Bonche-Koonin-Negele (BKN) energy density functional \cite{BKN76}.
This functional assumes the spin-isospin symmetry without 
the spin-orbit interaction.
Therefore, each orbit has the four-fold degeneracy.
We adopt the Cartesian grid representation of square mesh
of 1.1 fm
in the rectangular box \cite{NY05}.
The model space is set to be $12\times 12\times 18$ fm$^3$
for the reaction $^{16}$O$+\alpha\rightarrow^{20}$Ne
and $^{12}$C$+\alpha\rightarrow^{16}$O,
$12\times 12\times 24$ fm$^3$
for the system $^{16}$O$+^{16}$O$\rightarrow^{32}$S.
More details can be found in Refs.~\cite{WN16,WN17}.

\subsection{Collective reaction path}

It is significantly facilitated by an approximation to sacrifice
the full self-consistency among Eqs. (\ref{eq:mfHFB}),
(\ref{eq:ASCC1-2}), and (\ref{eq:ASCC2-2}).
Instead, we adopt $\Qhat(q+\delta q)\approx \hat{Q}(q)$
for Eq. (\ref{eq:mfHFB}).
Since $\hat{Q}(q)$ is a smooth function of $q$,
this is reasonable for a small step size $\delta q$.
With this approximation,
the Hamiltonian $\Hhat_M(q+\delta q)$ is now given by
$\Hhat_M(q+\delta q)\approx \Hhat_M(q)=\Hhat-\lambda(q) \hat{Q}(q)$.
The Lagrange multiplier $\lambda(q)$ is determined by the constraint on
the step size,
\begin{equation}
 \bra{\phi(q+\delta q)} \hat{Q}(q) \ket{\phi(q+\delta q)}
 = \delta q .
\label{step_size_constraint}
\end{equation}
We use the imaginary-time method for solution of Eq. (\ref{eq:mfHFB}).
In this way, the system moves from $\ket{\phi(q)}$ to $\ket{\phi(q+\delta q)}$,
obtaining a new state $\ket{\phi(q+\delta q)}$ on the collective path.

The finite amplitude method \cite{NIY07,INY09,AN11,Sto11,AN13}
is used for solution of
Eqs. (\ref{eq:ASCC1-2}) and (\ref{eq:ASCC2-2}).
Solving Eqs. (\ref{eq:ASCC1-2}) and (\ref{eq:ASCC2-2})
at $\ket{\phi(q+\delta q)}$,
the generators are updated from
$\Qhat(q)$ to $\Qhat(q+\delta q)$.
Then, we construct the next state, $\ket{\phi(q+2 \delta q)}$,
by solving Eq. (\ref{eq:mfHFB}) with the Hamiltonian
$\Hhat_M(q+2\delta q)\approx \Hhat_M(q+\delta q)
=\Hhat -\lambda(q+\delta q)\Qhat(q+\delta q)$.
Continuing this iteration, we will obtain a series of states,
$\ket{\phi(q=0)}$,
$\ket{\phi(\delta q)}$,
$\ket{\phi(2\delta q)}$,
$\ket{\phi(3\delta q)},\cdots$, forming a collective path.
The solution of collective path converges when $|\delta q|$ is small,
which justifies the adopted approximation,
$\Qhat(q+\delta q)\approx \Qhat(q)$.

The initial state to start this procedure can be chosen as any
state satisfying Eqs. (\ref{eq:mfHFB}), (\ref{eq:ASCC1-2}), and
(\ref{eq:ASCC2-2}).
There are some states which obviously satisfy these conditions;
a stationary state of the fused system
(Hartree-Fock ground and meta-stable states),
and the state of two well separated nuclei before collision.
We obtain many solutions of
Eqs. (\ref{eq:ASCC1-2}), and (\ref{eq:ASCC2-2}),
among which we choose the $\Qhat(q)$ and $\Phat(q)$ of
the lowest mode of excitation at these initial states.
Then, at $\ket{\phi(q+\delta q)}$, we choose
$\Qhat(q+\delta q)$ and $\Phat(q+\delta q)$
close to $\Qhat(q)$ and $\Phat(q)$.
In other words,
we assume a smooth and continuous reaction path.

\subsection{Mapping to the relative distance between projectile and target}

Since the scale of the collective coordinate $q$ is arbitrary,
the inertial mass $M(q)$ with respect to $q$ is also arbitrary.
In the asymptotic region where the two colliding nuclei
are well apart, it is natural to
adopt the relative distance $r$ between projectile and target
as the collective coordinate.
The definition of the relative distance becomes ambiguous after the two
nuclei touch.
However, this does not matter in the present approach.
As far as the one-to-one correspondence between $q$ and $r$
is guaranteed, the collective Hamiltonian in the coordinate $r$
\begin{equation}
\tilde{H}_c(r,p_r)=\frac{1}{2}\tilde{B}(r) p_r^2 + \tilde{V}(r)
\label{collective_Hamiltonian_r}
\end{equation}
is equivalent to Eq. (\ref{collective_Hamiltonian}),
whatever the definition of $r$ is.
Here, the potential is given as $\tilde{V}(r)=V(q(r))$ and
the momentum as $p_r=p \cdot dq/dr$.
The inertial mass should be transformed to
\begin{eqnarray}
\tilde{B}(r)=B(q)\left(\frac{dr}{dq}\right)^{2} .
\label{mass}
\end{eqnarray}
Here, the derivative $dr/dq$ can be obtained by use of
the local generator $\hat{P}(q)$ as,
\begin{equation}
\frac{dr}{dq} =
\frac{d}{dq}\langle \phi(q) |\hat{r}| \phi(q) \rangle
 =\langle \phi(q) | [\hat{r},\frac{1}{i}\hat{P}(q) ] | \phi(q) \rangle
.
\label{mass3}
\end{equation}
As a merit of this coordinate transformation from
the original $q$ to the intuitively chosen $r$,
we may obtain a intuitive physical picture of the collective dynamics.
Nevertheless, the dynamics are equivalent.

Equation (\ref{collective_Hamiltonian_r})
looks at first glance that we do not need to find the collective
submanifold (the coordinate $q$).
However, to achieve the proper inertial mass and the potential,
we need to know $B(q)$ and $V(q)$ calculated on $\ket{\phi(q)}$.
We should note the difference between ``mapping to'' and ``assuming'' the
collective coordinate:
$\tilde{V}(r)$ is different from the potential obtained with
the constrained minimization for $\bra{\phi}\hat{r}\ket{\phi}=r$.
$\tilde{B}(r)$ is not the inertia along the coordinate direction of $r$.

\section{Sub-barrier fusion reaction}
\label{sec:result}

\subsection{$^{16}$O$+\alpha\rightarrow ^{20}$Ne and
$^{16}$O$+^{16}$O$\rightarrow ^{32}$S}

For these reactions, the results are shown in Ref.~\cite{WN17}.
We recapitulate the result in this subsection.

For the $^{16}$O$+\alpha$ system,
we successfully obtain the reaction path to connect the ground state
of $^{20}$Ne and $^{16}$O$+\alpha$ asymptotically.
The ground state of $^{20}$Ne is calculated as a deformed nucleus with
a prolate shape.
The lowest mode of excitation corresponds to the $K=0$ octupole vibration.
According to the ASCC method, we calculate the reaction path which
smoothly ends up at $^{16}$O$+\alpha$.
This smooth reaction path cannot be obtained by the
Hartree-Fock calculation with the mass quadrupole operator
as a constraint operator.

The calculated inertial mass for the relative distance $r$,
$M(r)=1/\tilde{B}(r)$, exactly matches the reduced mass ($3.2 m_N$)
for $^{16}$O$+\alpha$ at the limit of $r\rightarrow\infty$,
and increases as $r$ decreases.
Near the ground state of $^{20}$Ne, 
$M(r)$ shows a peculiar rising,
which indicates that the reaction path becomes almost orthogonal
to the relative distance $r$.
See Eq. (\ref{mass}).
With $dr/dq\rightarrow 0$, we have $\tilde{B}(r)\rightarrow 0$ and
$M(r) \rightarrow \infty$.

For the $^{16}$O$+^{16}$O case,
the reaction path leads to a meta-stable state in $^{32}$S,
namely the prolate superdeformed local minimum.
We tried to extend the path toward a more compact shape,
however, we could not reach the ground state of $^{32}$S.
In the present calculation, the ground state of $^{32}$S has
a triaxial shape with $\gamma=36^\circ$.

The calculated inertial mass for the relative distance $r$,
again, exactly matches the reduced mass ($8 m_N$)
for $^{16}$O$+^{16}$O at $r\rightarrow\infty$.
Then, $M(r)$ increases as $r$ decreases after the touching.
$M(r)$ reaches about twice of the reduced mass near the
superdeformed local minimum in $^{32}$S.
Beyond this point, it shows a rapid increase and
the collective reaction path stops.

We requantize the collective Hamiltonian of
Eq. (\ref{collective_Hamiltonian}),
or equivalently, (\ref{collective_Hamiltonian_r}),
then, calculate the fusion cross section at sub-barrier energies.
The result indicates the importance of the dynamical inertial mass.
Compared to the calculation with a constant reduced mass,
the one with $M(r)$ produces smaller fusion cross section
at deep sub-barrier energies.
Especially, for the $^{16}$O$+^{16}$O case,
the reduction factor could be two order of magnitude at very low energy.

\subsection{$^{12}$C$+\alpha$}

In this subsection, we present somewhat preliminary results for
the reaction of $^{12}$C$+\alpha$.
The ground state of $^{12}$C has an oblate shape.
Thus, even if we consider only the central collision,
there are orientation degrees of freedom to choose for this reaction.
Here, we consider the case with the axial symmetry, namely,
the alpha particle colliding head-on to the center of the disk ($^{12}$C).

At the spherical ground state of $^{16}$O, the lowest mode of excitation
corresponds to the octupole vibrational mode.
There are seven degenerate modes with different magnetic quantum numbers
$K$ of $Y_{3K}$.
We choose the $K=0$ mode to keep the axial symmetry.
Then, we may construct the ASCC collective path up to
the state with the octupole moment
$Q_{30}(q)\equiv \bra{\phi(q)}r^3Y_{30} \ket{\phi(q)}\approx 150$ fm$^3$.
However, the frequency $\omega^2(q)=B(q)C(q)$ quickly increases near
this ending point and we cannot construct the path beyond that.
We also try to find the collective path from the opposite limit,
namely well-apart $^{12}$C and $\alpha$.
In this case, we obtain the self-consistent path up to
$Q_{30}(q)\approx 300$ fm$^3$.
However, we find difficulties to select the proper generator $\Qhat(q)$
beyond this point toward smaller $Q_{30}$.
In fact, beyond these ending points,
the state becomes triaxial and we cannot keep the axial symmetry
satisfied in a region of $150\lesssim Q_{30} \lesssim 300$ fm$^3$.

\begin{figure}[tb]
\begin{minipage}[b]{0.55\textwidth}
\includegraphics[width=\textwidth]{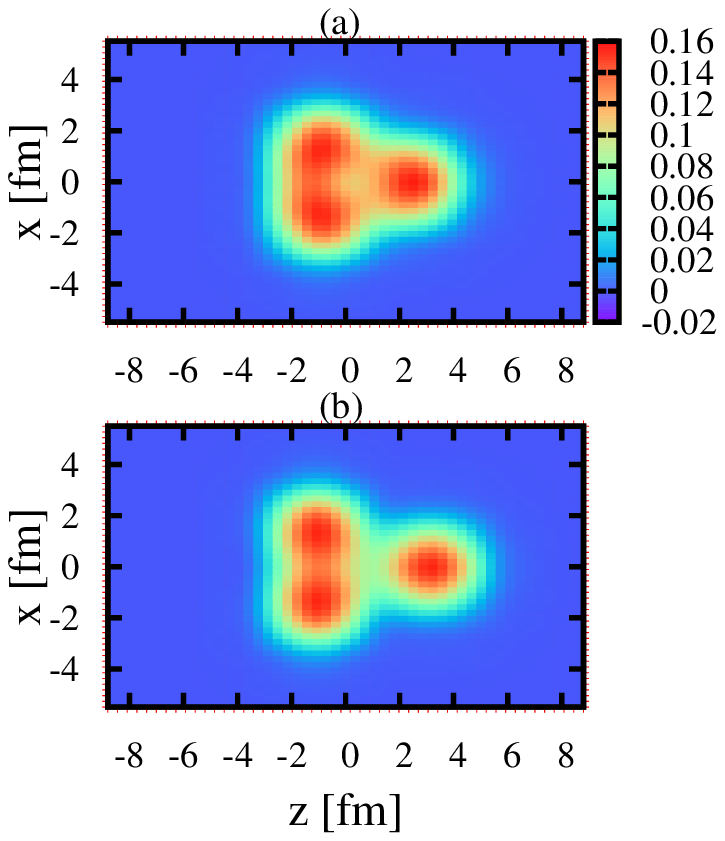}
\caption{
Density distribution in the $y=0$ plane on the collective path
for $^{12}$C$+\alpha \rightarrow^{16}$O, at
(a) $Q_{30} =150$ fm$^{3}$, and 
(b) $Q_{30} = 300$ fm$^{3}$.
}
\label{fig:12C_density}
\end{minipage}
\hfill
\begin{minipage}[b]{0.4\textwidth}
\includegraphics[width=\textwidth]{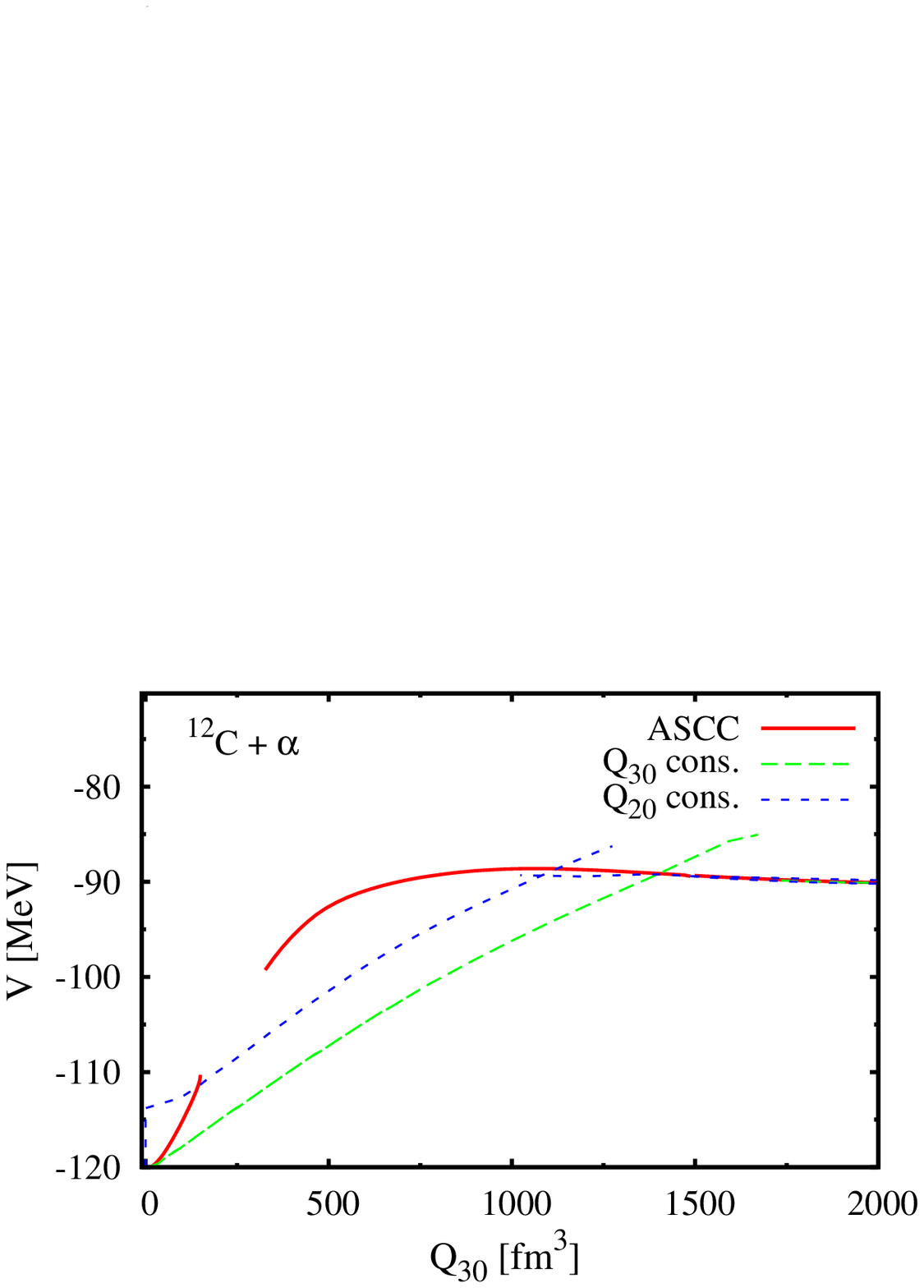}
\includegraphics[width=\textwidth]{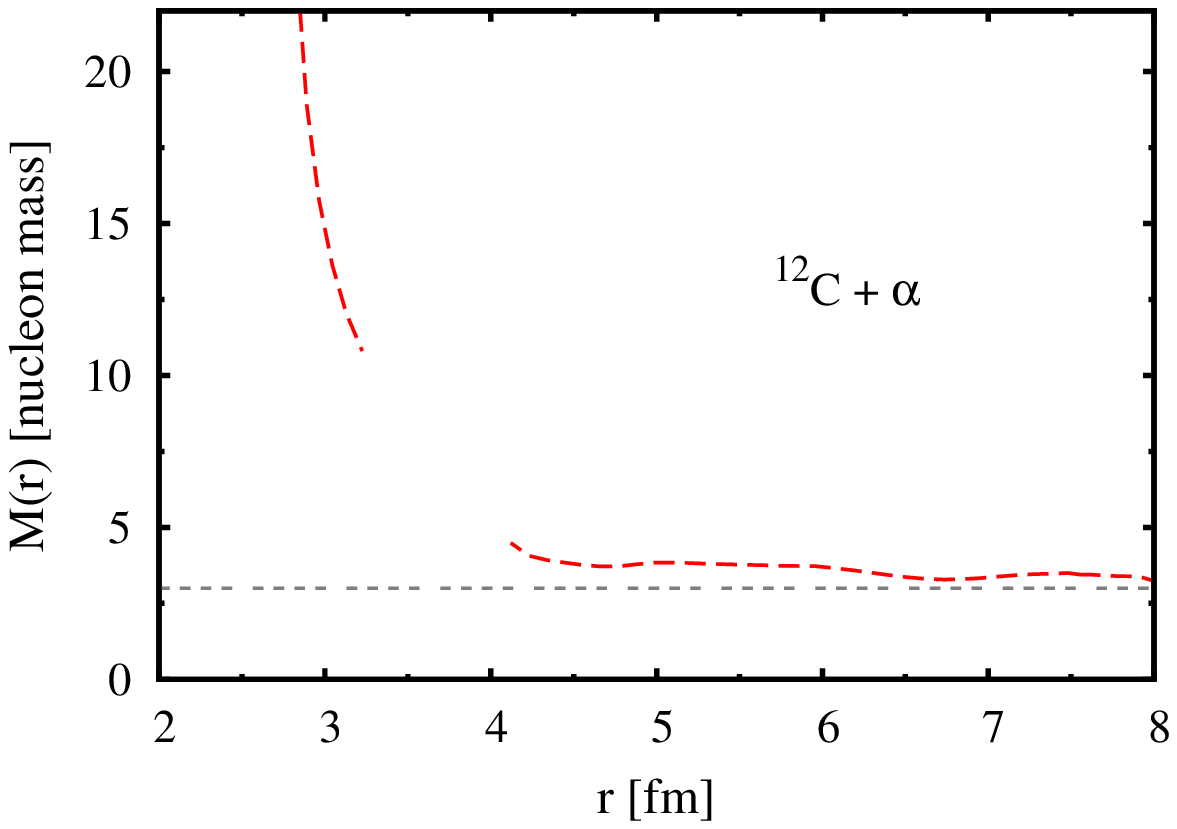}
\caption{
Top panel: Potential energy as a function of the octupole
moment $Q_{30}$ for $^{12}$C$+\alpha \rightarrow^{16}$O.
The red solid line represents the result of the ASCC method, while
the blue (green) dashed line indicates that with the CHF
calculation with $Q_{20}$ ($Q_{30}$) as a constraint.
Bottom panel: Inertial mass with respect to the
relative distance $r$, as a function of $r$.
}
\label{fig:12C_potential_mass}
\end{minipage}

\end{figure}

The density distribution profiles near these two ending points are
shown in Fig.~\ref{fig:12C_density}.
We may recognize a structure of $^{12}$C$+\alpha$ in both panels
(a) and (b).
Between these two states, the $^{12}$C nucleus tries to rotate
to break the axial symmetry.
This may be natural because the $\alpha$ particle is able to touch
the $^{12}$C nucleus at larger $r$,
if we tilt the $^{12}$C nucleus.
Currently, it is not clear to us whether there is a reaction path
connecting these two ending points.

The obtained potential and inertial mass are presented in
Fig.~\ref{fig:12C_potential_mass}.
In the top panel of Fig.~\ref{fig:12C_potential_mass}, we also
show the results of the constrained Hartree-Fock (CHF) calculation with
the mass quadrupole $Q_{20}$ and octupole $Q_{30}$ constraints,
for comparison.
The $Q_{20}$-constrained calculation produces no octupole deformation
near the ground state of $^{16}$O, then at a certain point, it starts
to produce the octupole deformation.
The ASCC calculation produces the smooth potential curve
from the separated $^{12}$C$+\alpha$ toward the touching region.
On the other hand, the CHF calculations produces
discontinuous potential curves in the region of
$1,000\lesssim Q_{30} \lesssim 1,500$ fm$^3$.
This indicates a clear advantage of the ASCC method over the
conventional constrained calculations.

The calculated inertial mass shows a behavior similar to previous cases
of $^{16}$O$+\alpha$
(bottom panel of Fig.~\ref{fig:12C_potential_mass}).
Near the ground state of the fused system $^{16}$O,
the inertial mass rapidly increases.
This suggests the property of the collective coordinate,
that $q$ is approximately orthogonal to $r$ near the ground state.
In the asymptotic region of $r\rightarrow\infty$,
$M(r)$ is identical to the reduced mass, $3 m_N$.
This is a desired feature because the relative distance $r$ should be
a collective coordinate in the asymptotic region,
then, the inertial mass should be equal to the reduced mass.

\section{Summary}
\label{sec:summary}

We have investigated the fusion reaction dynamics at sub-barrier energies,
with the adiabatic self-consistent coordinate (ASCC) method
using the BKN energy density functional.
The numerical calculation is based on the imaginary-time method
and the finite amplitude method in the three-dimensional Cartesian mesh
representation.
We have successfully constructed the reaction path for the
sub-barrier fusion reactions for $^{16}$O$+\alpha$ and $^{16}$O$+^{16}$O,
keeping the axial symmetry.
On the other hand, the reaction path for $^{12}$C$+\alpha$ seems to be
significantly deviated from the axial symmetric configuration after
two nuclei touch.
We have not succeeded to obtain the self-consistent reaction path
for this system, so far.

The obtained reaction paths are always continuous.
This is in contrast to the constrained Hartree-Fock calculation with
a fixed constraint operator.
The inertial mass always reproduces the reduced mass in the
asymptotic region ($r\rightarrow\infty$).
The ASCC mass corresponds to the Thouless-Valatin mass at
the Hartree-Fock minima, which is known to reproduce the exact
mass for the center-of-mass motion \cite{RS80}.
Our result suggests the collective motion is very different from the
simple relative motion, after the touching point.
Therefore, the inertial masses with respect to the relative
distance $r$ show a rapid increase,
which leads to a hindrance of the sub-barrier fusion cross sections.

This work is supported in part by JSPS KAKENHI Grant No. 25287065,
by Interdisciplinary Computational Science Program in CCS,
University of Tsukuba, and
by JSPS and NSFC under the Japan-China Scientific Cooperation Program.

%


\begin{thebibliography}{9}
\bibitem{Sim12}
C. Simenel, 
Eur. Phys. J. A {\bf 48}, 152 (2012).
\bibitem{NMMY16}
T. Nakatsukasa, K. Matsuyanagi, M. Matsuo, and K. Yabana,
Rev. Mod. Phys. {\bf 88}, 045004 (2016).
\bibitem{SY16}
K. Sekizawa and K. Yabana,
Phys. Rev. C {\bf 93}, 054616 (2016).
\bibitem{UOS16}
A. S. Umar, V. E. Oberacker, and C. Simenel,
Phys. Rev. C {\bf 94}, 024605 (2016).
\bibitem{BMRS16}
A. Bulgac, P. Magierski, K. J. Roche, and I. Stetcu,
Phys. Rev. Lett. {\bf 116}, 122504 (2016).
\bibitem{MSW17}
P. Magierski, K. Sekizawa, and G. Wlazłowski,
Phys. Rev. Lett. {\bf 119}, 042501 (2017).
\bibitem{Sek17}
K. Sekizawa, Phys. Rev. C {\bf 96}, 014615 (2017).
\bibitem{USY17}
A. S. Umar, C. Simenel, and W. Ye,
Phys. Rev. C {\bf 96}, 024625 (2017).
\bibitem{Neg82}
J. Negele,
Rev. Mod. Phys. {\bf 54}, 913 (1982).
\bibitem{MNM00}
M. Matsuo, T. Nakatsukasa, and K. Matsuyanagi,
Prog. Theor. Phys. {\bf 103}, 959 (2000).
\bibitem{HNMM07}
N. Hinohara, T. Nakatsukasa, M. Matsuo, and K. Matsuyanagi,
Prog. Theor. Phys. {\bf 117}, 451 (2007).
\bibitem{HNMM09}
N. Hinohara, T. Nakatsukasa, M. Matsuo, and K. Matsuyanagi,
Phys. Rev. C {\bf 80}, 014305 (2009).
\bibitem{Nak12}
T. Nakatsukasa, 
Prog. Theor. Exp. Phys. {\bf 2012}, 01A207 (2012).
\bibitem{RS80}
P. Ring and P. Schuck,
The Nuclear Many-Body Problem, Springer-Verlag, New York, (1980).
\bibitem{AER01}
M. Anguiano, J. Egido, and L. Robledo,
Nucl. Phys. A {\bf 696}, 467 (2001).
\bibitem{DSNR07}
J. Dobaczewski, M. V. Stoitsov, W. Nazarewicz, and P.-G. Reinhard,
Phys. Rev. C {\bf 76}, 054315 (2007).
\bibitem{DBBLL09}
T. Duguet, M. Bender, K. Bennaceur, D. Lacroix, and T. Lesinski,
Phys. Rev. C {\bf 79}, 044320 (2009).
\bibitem{PT62}
R. E. Peierls and D. J. Thouless, 1962, Nucl. Phys. 38, 154.
\bibitem{MMSK80}
T. Marumori, T. Maskawa, F. Sakata, and A. Kuriyama,
Prog. Theor. Phys. {\bf 64}, 1294 (1980).
\bibitem{YK87}
M. Yamamura and A. Kuriyama,
Prog. Theor. Phys. Suppl. {\bf 93}, 1 (1987).
\bibitem{BKN76}
P. Bonche, S. Koonin, and J.~W. Negele,
Phys. Rev. C {\bf 13}, 1226 (1976).
\bibitem{NY05}
T. Nakatsukasa and K. Yabana,
Phys. Rev. C {\bf 71}, 024301 (2005).
\bibitem{WN16}
K. Wen and T. Nakatsukasa,
Phys. Rev. C {\bf 94}, 054618 (2016).
\bibitem{WN17}
K. Wen and T. Nakatsukasa,
Phys. Rev. C {\bf 96}, 014610 (2017).
\bibitem{NIY07}
T. Nakatsukasa, T. Inakura, and K. Yabana,
Phys. Rev. C {\bf 76}, 024318 (2007).
\bibitem{INY09}
T. Inakura, T. Nakatsukasa, and K. Yabana,
Phys. Rev. C {\bf 80}, 044301 (2009).
\bibitem{AN11}
P. Avogadro and T. Nakatsukasa,
Phys. Rev. C {\bf 84}, 014314 (2011).
\bibitem{Sto11}
M. Stoitsov, M. Kortelainen, T. Nakatsukasa, C. Losa, and W. Nazarewicz,
Phys. Rev. C {\bf 84}, 041305 (2011).
\bibitem{AN13}
P. Avogadro and T. Nakatsukasa,
Phys. Rev. C {\bf 87}, 014331 (2013).
%
\end{thebibliography}
\end{document}